# Drag coefficient for irregularly shaped grains: rotational dependence at high Reynolds numbers

**Álvaro Vergara**[1], **Deheng Wei**[1,†] **and Raúl Fuentes**[1]

[1]Institute of Geomechanics and Underground Technology, RWTH Aachen, Aachen 52074, Germany

The nature and behaviour of the drag coefficient $C_D$ of irregularly shaped grains within a wide range of Reynolds numbers $Re$ is discussed. Using computational fluid dynamics (CFD) tools, the behaviour of the boundary layer at high $Re$ has been determined by applying the Reynolds Averaged Navier-Stokes turbulence model (RANS). The dependence of the mesh size and the grid resolution in the modelling are validated with the previous experimental results applied in flow around isolated smooth spheres. The drag coefficient for irregularly shaped grains is shown to be higher than that for spherical shapes, also showing a strong drop in its value at high $Re$ (drag crisis) but lower than that of the sphere. The influence of the angle of incidence of the flow with respect to the particle is analysed, where our findings show an interesting oscillatory behaviour of the drag coefficient as a function of the angle of incidence, fitting the results to a sine-squared interpolation, predicted for particles within the Stokes' laminar regime ($Re \ll 1$) and for bodies with an ellipsoidal shape (elongated and flattened spheroids) up to $Re = 2000$. The statistical analysis shows a Weibullian behaviour of the drag coefficient when random polar and azimuthal rotation angles are considered.

† Email address for correspondence: wei@gut.rwth-aachen.de

## 1. Introduction

A variety of physical phenomena involve irregularly shaped particles. Geosciences, astrophysics, the chemical industry, mineral treatment and material sciences are some of the disciplines that encompass the study of the interactions of particulate systems with fluid flow. Segregation of granular matter, asteroid penetration into the atmosphere, sediment transport in natural systems, handling of mineral grains in the mining industry, are some engineering applications where this is particularly relevant. This type of system is usually studied from a hydrodynamic perspective using the forces acting on the surface of the body.

This involves analysing the drag coefficient $C_D = |\mathbf{F}_D| / \left(\frac{1}{2}\rho_f |\mathbf{u}_\infty|^2 A_p\right)$, as a dimensionless quantity capable of describing the resistance force exerted by the body within a fluid medium, where $\mathbf{F}_D$ is the drag force exerted on the body, $\rho_f$ the density of the fluid, $\mathbf{u}_\infty$ the terminal or free stream velocity of the fluid and $A_p$ the projected area of the body on a normal plane to the direction of flow.

Drag coefficient is used to optimise aerodynamic designs of solid bodies, or to describe the existing forces of particles immersed in a fluid medium. Its engineering applications extend to aerodynamic design of vehicles such as airplanes, automobiles, and ships (Bayindirli & Çelik 2018), net design in aquaculture and oceanography (Tsukrov et al. 2011), optimisation of pneumatic transport in food industry (Defraeye et al. 2013), structure design (Hamelin et al. 2013), hydrodynamic behaviour of plant layers (Etminan et al. 2017) and sports (Beratlis et al. 2019), sedimentation and transport of sediments, among other applications.

However, in other disciplines more interest is usually paid to the dynamics aspects of drag in discrete bodies. Regarding the treatment of individual geometries, the drag coefficient has been well studied and understood for spherical geometries (Achenbach 1972; Constantinescu & Squires 2004; Haider 1989; M. Sun et al. 2005) as a basis for extending it to more complex geometries, from ellipsoids (Ouchene et al. 2015; Sanjeevi et al. 2018; Sanjeevi & Padding 2017), spherical clusters (Tran-Cong et al. 2004), cylinders (Lam & Lin 2009) and agglomerates (Chen et al. 2022), to irregularly-shaped particles (Sommerfeld & Qadir 2018; Q. Sun et al. 2018; Wang et al. 2018; Zhang et al. 2023) and granular media (Guillard et al. 2014; Jing et al. 2022). Michaelides & Feng (2023) recently performed a complete review of the different models to estimate the drag coefficients for both symmetrical particles and irregularly-shaped particles.

In general, the drag coefficient depends on the Reynolds number, $Re$, especially for Reynolds numbers below $10^4$, above this value, the drag coefficient remains practically constant due to turbulence. Additionally, for smooth bodies (such as spheres and cylinders) in highly turbulent flows there is also a drag crisis, where the drag coefficient decreases drastically up to a critical value, beyond which it then increases gradually with increasing $Re$ until reaching a relatively constant value. This drag crisis is related to the turbulent transition of the boundary layer and the effects of pressure patterns on viscous forces when $Re$ is relatively large. However, its existence and the behaviour during this transition has not been shown for irregularly shaped particles, despite initial evidence showing that.

The irregularity of the particle surface directly impacts the drag coefficient. Zhang et al. (2023) proposed various drag coefficient models for irregularly shaped particles for $Re \leq 200$. The authors determined that $C_D$ increases as the particle surface becomes more irregular. They also showed that $C_D$ varies slowly at low Reynolds numbers for different fluid incidence angles, while it does so more rapidly at high $Re$. However, they did not study the behaviour for very large $Re$.

The effect of rotational dependence of smooth bodies on drag is well described by Sanjeevi & Padding, (2017), who derived a periodic behaviour of the drag coefficient as the ellipsoidal particle is rotated under the circumstance where its longest axis, shortest, and flow directions are always on the same plane, for $Re$ up to 2000. However, many questions about the influence of surface irregularity and the complete randomness of rotation on the fluid dynamic behaviour of the particle still remain to be resolved, especially again under turbulent regimes.

The objective of this work is to characterise the fluid dynamic behaviour (drag coefficient) of irregularly-shaped particles in a wide spectrum of flow regimes, $Re = [1 - 10^7]$, and study the influence of rotation angles. This article is divided into the following sections. In section 2 we describe the computational fluid dynamics (CFD) tools applied to resolve pressure and velocity fields, within the setup geometry, both for model validation and for obtaining the drag forces for irregular geometries. The results of the drag and the rotational influence for the range of laminar and turbulent flow are presented in section 3, highlighting and discussing the different findings of the fluid dynamic behaviour of the drag for this type of geometries. The main conclusions of this research are summarised in section 4.

## 2. Methods

### 2.1. *Numerical resolution and geometric considerations*

The flow development is computed by numerically solving the Navier-Stokes equations for viscous, incompressible and isothermal flow:

$$
\begin{aligned}
&\boldsymbol{\nabla} \cdot \mathbf{u} = 0 \\
&\rho_f (\mathbf{u} \cdot \boldsymbol{\nabla}) \mathbf{u} = \boldsymbol{\nabla} \cdot \left[ -pI + \mu_f (\nabla \mathbf{u} + (\nabla \mathbf{u})^T) \right]
\end{aligned}
\tag{2.1}
$$

where $\mathbf{u}$ is the velocity field, $\rho_f$ the fluid density, $\mu_f$ the fluid dynamic viscosity and $p$ the pressure field in the system.

The flow regime at high $Re$ is evaluated using the Reynolds Averaged Navier-Stokes turbulence model (RANS) applied to the transport equations for shear stress (SST k-ω) viscous model, due to its high capacity to resolve turbulent elements in the boundary layer, improving its treatment at high Reynolds numbers (Menter 1994). This numerical resolution is obtained from the pressure-velocity coupled scheme (based on pressure), which allows the simultaneous resolution of the equations of motion and continuity, ensuring greater robustness of the numerical solution. The spatial discretization follows a second order method, given the irregular flow conditions in the boundary layer, especially in the turbulent regime.

The irregular geometry of the grain follows the model of Wei et al. 2018 who proposed a practical method for generating irregularly-shaped particles using a fractal dimension as a descriptor parameter of the morphology. It uses spherical harmonics that allow to define and control the morphology of the particle from its fractal properties, both its shape, its roundness and texture, based on the model presented by Wei et al. 2018. The generated irregular particle used in this paper is shown in FIGURE 1($a$,$b$).

The particle is characterised by its axial asymmetry caused by the presence of dimples and ridges randomly distributed over its entire surface. Its fractal dimension, $D_f$, is 2.287 and its characteristic lengths in the longest, intermediate and shortest dimensions of a box containing the particle inside are 2.87 mm, 2.09 mm and 1.27 mm respectively.

We define the Reynolds number based on the equivalent particle diameter, $d_p$, as $Re = \rho_f |\mathbf{u}_\infty| d_p / \mu_f$, with $\mathbf{u}_\infty$ the free stream velocity field in the direction of the flow. In the next section we show the influence of the different definitions of the equivalent diameter on the drag coefficient, justifying our choice in the definition of $Re$.

### 2.2. *Numerical setup and grid resolution influence*

The configuration applied in the simulations emulates those developed in physical experimentation, with a solid body inside a wind tunnel through which the fluid flows and where the resulting forces on the surface of the body under study are analysed. FIGURE 2$a$ shows the geometry used in all the simulations. These dimensions were chosen to avoid any influence of the geometry walls. The boundary conditions consider an inlet that allows the constant flow of fluid at free stream velocity (velocity whose value allows controlling the Reynolds number), a pressure outlet past the irregularly-shaped grain, and a no-slip condition on the walls.

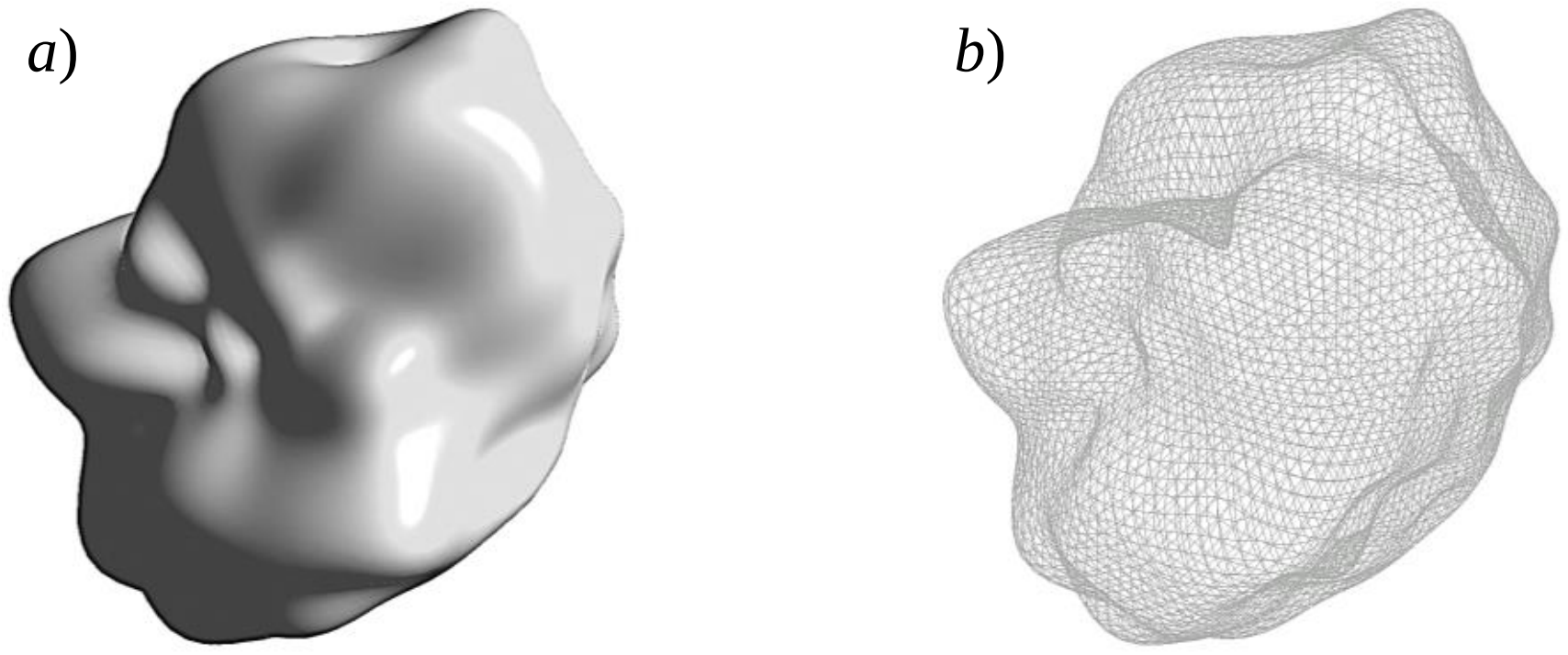

FIGURE 1. *a)* Irregularly shaped grain main geometry. *b)* Generic mesh triangulation used for flow resolution in CFD.

To determine the resolution of the modelling grid, calibration is performed using an isolated smooth sphere and comparing its drag coefficient as a function of Reynolds number with the experimental values obtained by Achenbach 1972. Critically, we concentrate on the size of the mesh around the particle in order to capture the boundary layer effects even at high Reynolds numbers. FIGURE 2*b* shows the meshing detail in the zone close to the boundary layer, where the thick black line represents the refinement required to capture the information at the boundary between the irregularly-shaped grain and the fluid (smaller mesh elements). FIGURE 3*a* shows the comparison of the numerical model and the experimental data applied to the smooth sphere, obtained by Achenbach 1972 and fitted by Turton & Levenspiel 1986. The presence and good fitting of the drag crisis is reported at $Re > 3 \times 10^5$. This result allows to determine the resolution of the mesh, where as $Re$ increases, the size of the cells of the mesh needs to be smaller to capture the turbulent behaviour in the boundary layer and the wake zone. The mesh ratio, $x/d_p$, is used, with $x$ being the characteristic length of each mesh element on the particle surface and the red dashed line in FIGURE 3*a* shows the maximum mesh size required to obtain the reasonable results. This translates, for the different simulations, into a range of element numbers from 700,000 (for low $Re$) to more than 3,000,000 mesh elements (for extremely high $Re$). FIGURE 3*b* shows the good correlation between the predicted and measured data for the drag coefficient when using this largest mesh size.

The analysis of the rotational dependence of the drag coefficient is performed considering two types of particle rotation respect to the fluid flow: polar rotation $\alpha$ (on the meridional x-y plane) and azimuthal rotation $\beta$ (on the y-z plane) as shown in FIGURE 2*c*. For the first analysis, the complete rotation of the polar angle is considered (from 0 to $2\pi$, varying the angle every $\pi/6$) following the rotation matrix $\mathcal{R}_1 = (\alpha, 0)$, while a statistical analysis is performed to consider the combined effect of polar and azimuthal rotations according to the ordered pair $\mathcal{R}_2 = (\alpha, \beta)$ with randomly distributed values.

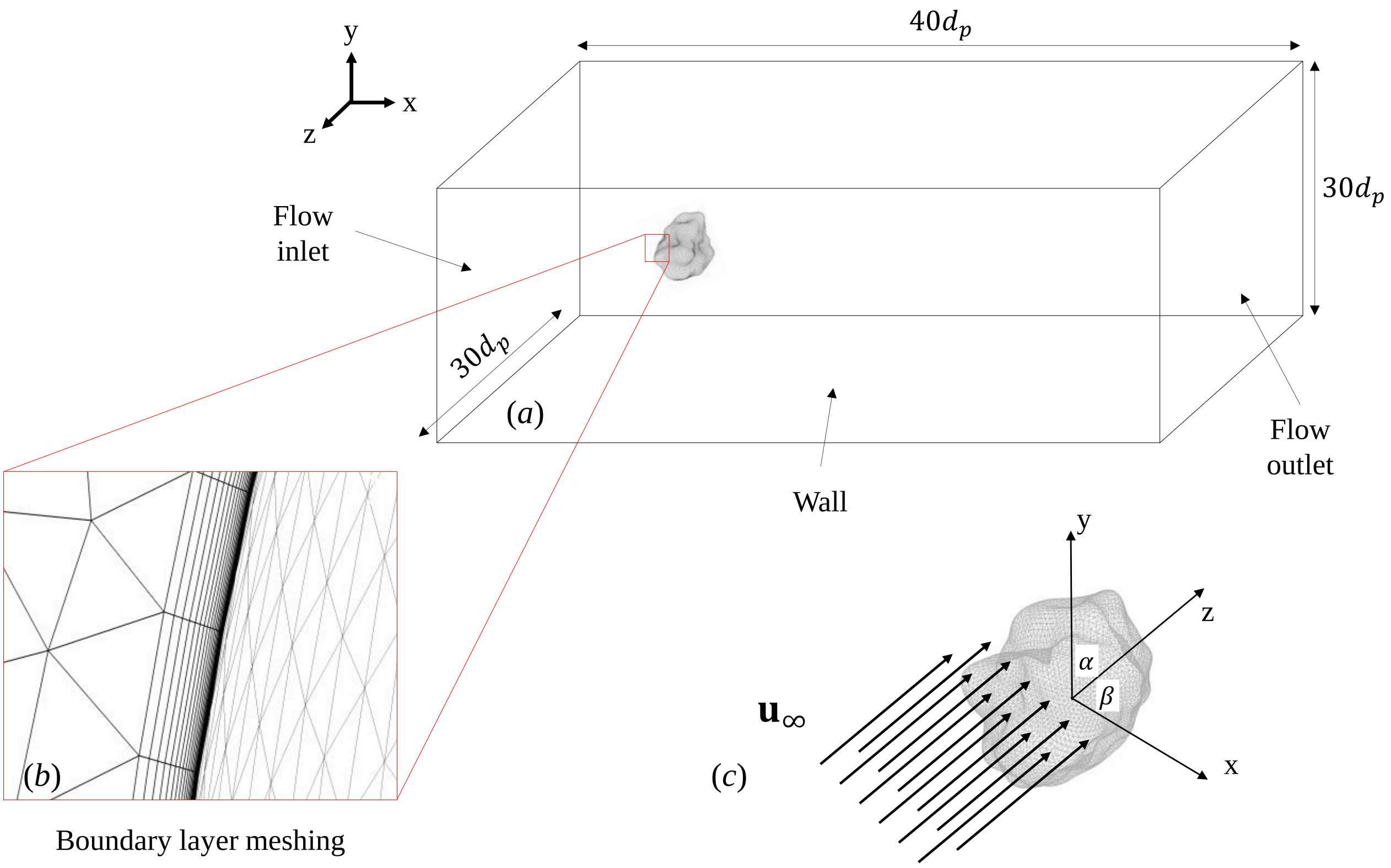

FIGURE 2. *a*) Computational domain used in the numerical study, whose dimensions make it possible to avoid the influence of the fluid-wall interaction on the fluid-particle interaction (not to scale). *b*) Unstructured mesh applied in conjunction with inflation methods to optimise turbulence capture in the boundary layer. *c*) Rotational consideration of polar and azimuthal angles ($\alpha$ and $\beta$ respectively) for orientational dependence analysis.

## 3. Results

### 3.1. *Effect of the Reynolds number*

FIGURE 4*a* shows the drag coefficient as a function of $Re$ at different flow incidence angles (polar rotation $\alpha$). In general, the drag coefficient decreases up to values of $Re \approx 10000$ (subcritical regime). Between these $Re$ values up to the drag crisis, $C_D$ is independent of the flow regime. Within this zone, the drag coefficient is usually greater than that of the sphere (solid black line in FIGURE 4($a$, $b$) for any incidence angle $\alpha$, despite having the same decreasing behaviour.

This shows the influence of the irregularity of the particle surface, altering the flow conditions inducing irregularities in the streamline patterns. As for smooth spheres, the irregular particle also suffers a drag crisis with an abrupt decrease of the drag coefficient for values of $Re \approx 100000$. In fact, as it is observed for dimpled and rough spheres, the irregular particle causes a more sudden drop of the drag coefficient, compared to the smooth sphere. This means that the roughness of the surface displaces the boundary layer, causing the drag coefficient to reach its critical value at lower Reynolds numbers. FIGURE 4*b* presents the drag crisis for the irregularly shaped particle, compared to the sphere. It can be observed that the minimum values of the drag coefficient are larger than those of the smooth sphere ($C_D \approx 0.07$). After this crisis, $C_D$ increases again until it is practically independent of the Reynolds number, but dependent on the incidence angle (see the colour scale). In summary, a common pattern of the existence of the four flow mechanisms zones in terms of $C_D$ arises: (1) subcritical, where $C_D$ decreases as $Re$ increases; (2) critical, where $C_D$ decreases sharply with growing $Re$; (3) supercritical, beyond the lowest value of $C_D$, where it increases slightly with increasing turbulence; and finally the (4) transcritical regime, where $C_D$ is practically constant at very high values of $Re$.

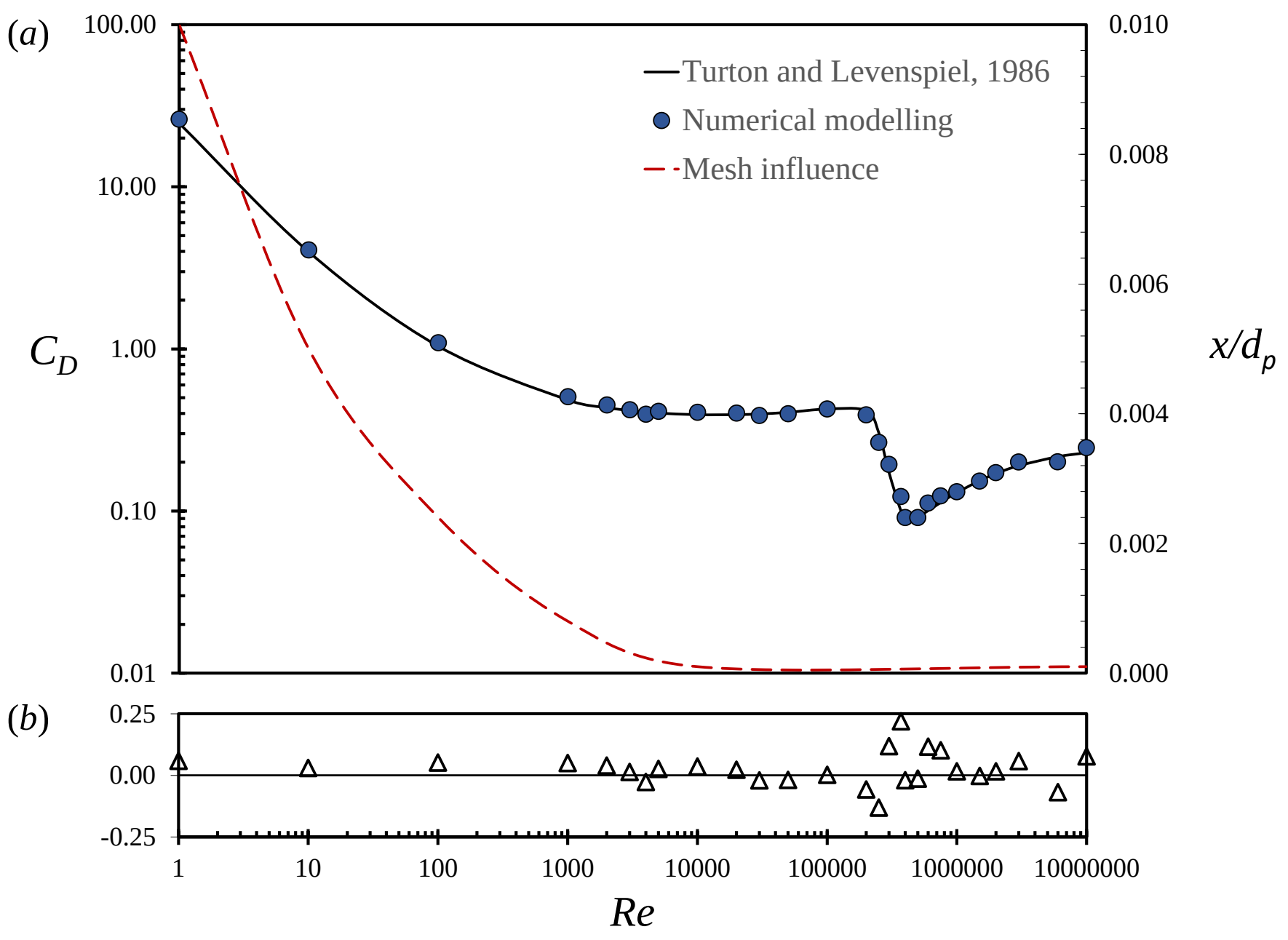

FIGURE 3. *a*) Drag coefficient $C_D$ as a function of the Reynolds number $Re$. The solid black line represents the fitted experimental values for the smooth sphere (Turton & Levenspiel 1986). The blue dots represent each numerical simulation. The red dashed line corresponds to the mesh resolution, showing a dependence on $Re$. *b*) Difference between experimental data and numerical results.

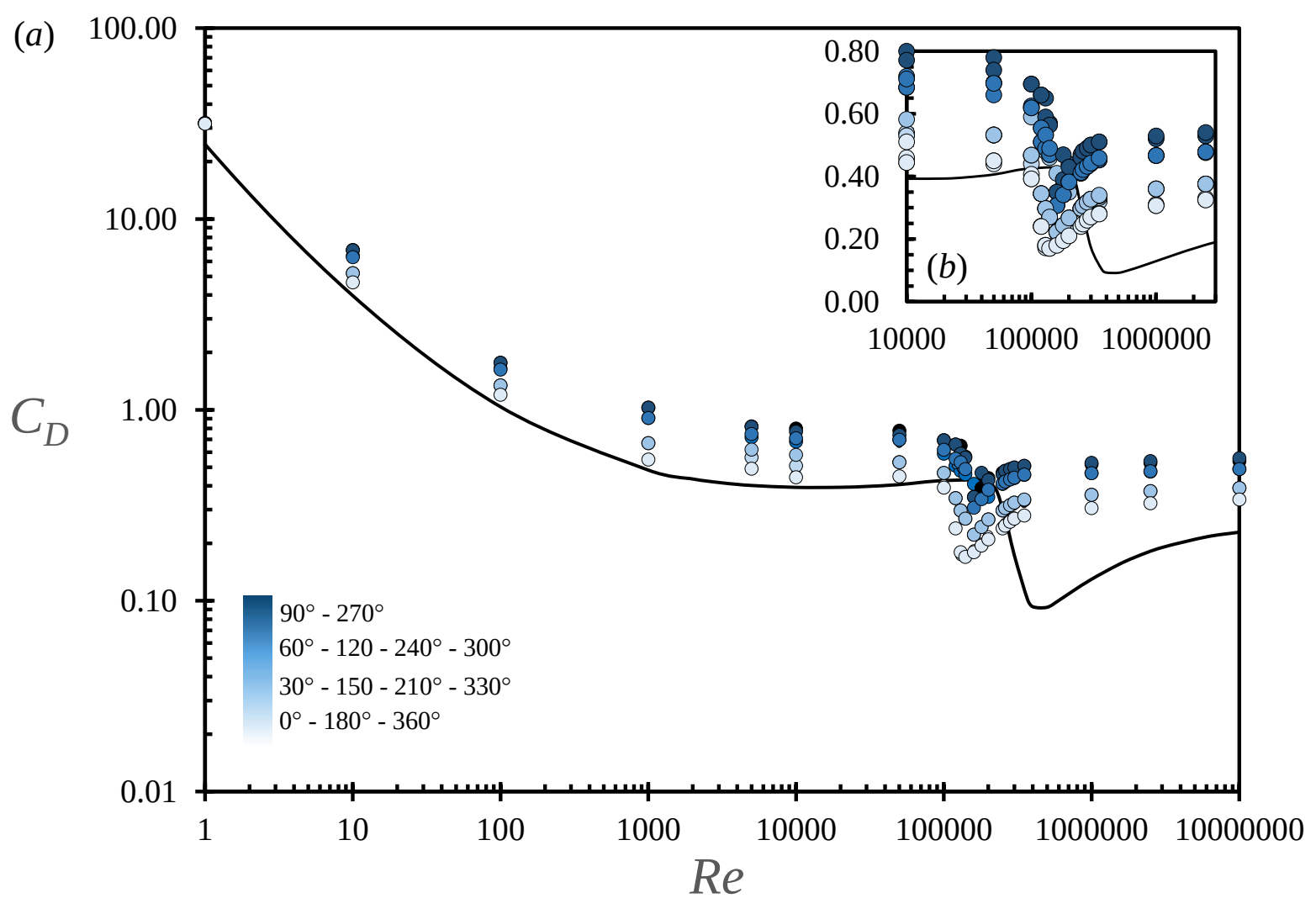

FIGURE 4. *a)* Drag coefficient $C_D$ as a function of the Reynolds number $Re$. The black solid line represents the values for the smooth sphere. The points represent the different simulations for the irregularly shaped particle. The colour scale shows the range of values of the polar rotation, $\alpha$. *b)* Drag coefficient $C_D$ as a function of the Reynolds number $Re$ in the drag crisis.

### 3.2. *Effect of the incidence angle*

FIGURE 5*a* shows that as the angle of rotation changes, the drag coefficient increases up to a maximum value. This behaviour reverses as the angle continues to change, repeating itself, similarly to a periodic pattern. This phenomenon, associated with the variation of the surface area directly exposed to the fluid flow, indicates a close relationship between the pressure patterns that influence the flow, even at high Reynolds numbers. To better elucidate this, a normalised drag coefficient, defined as $C_D^* \equiv \left(C_{D,a} - C_{D,min}\right)/\left(C_{D,max} - C_{D,min}\right)$ is derived to help understand the relationship between the variables. $C_{D,min}$ is equal to $C_{D,\alpha=0°}$ in our case, but in general, it would be the minimum value in the plot shown in FIGURE 5*a*, should another orientation be chosen as initial. $C_{D,max}$ represents the maximum value. FIGURE 5*b* presents $C_D^*$ as a function of the polar incidence angle $\alpha$, confirming the cyclic nature, but also, critically almost removing the effect of $Re$.

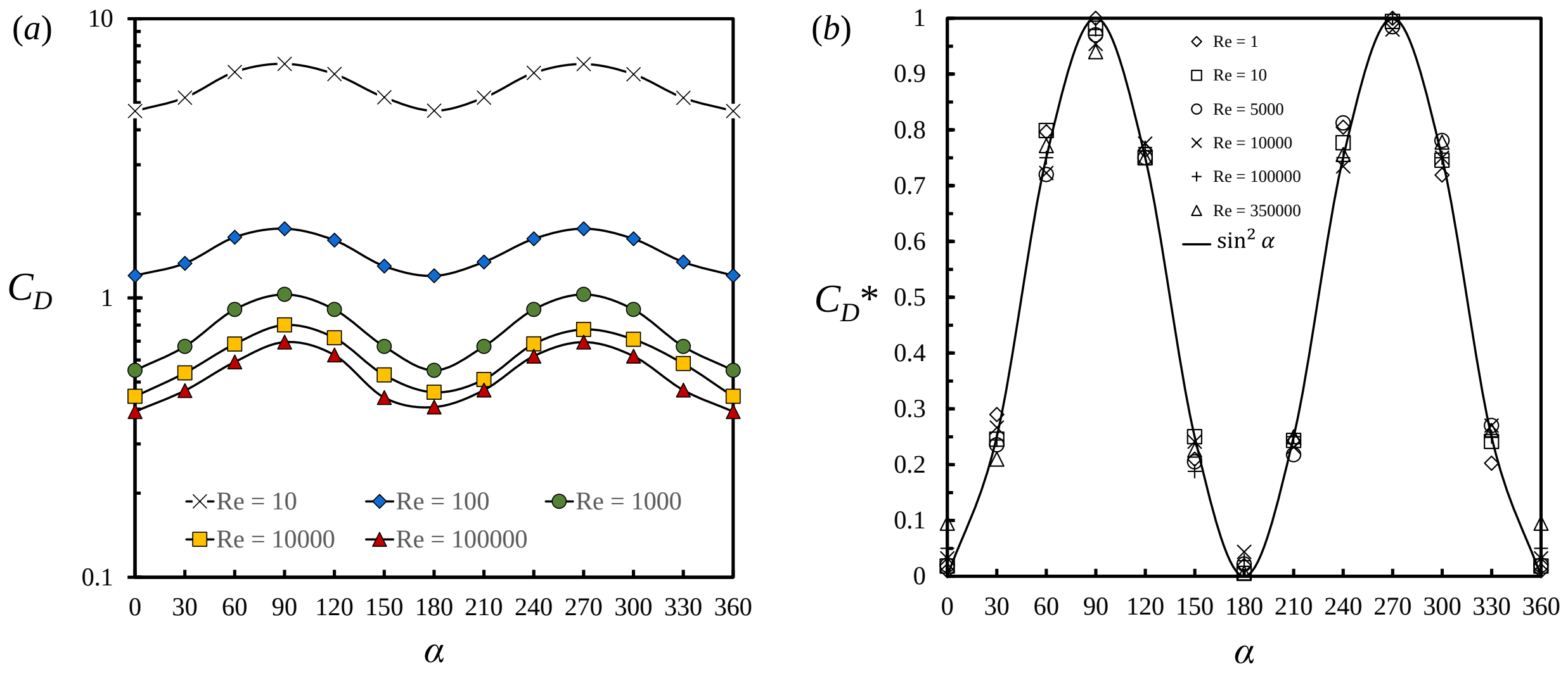

FIGURE 5. *a*) Drag coefficient $C_D$ as a function of the angle of incidence of the flow $\alpha$ at different $Re$. *b*) Normalised drag coefficient $C_D^*$ as a function of the angle of incidence of the flow $\alpha$ at different $Re$. The solid line represents the sine-squared function.

In fact, we observe that this normalised drag coefficient, is also represented by a sine-squared drag law (Equation 3.1) similarly to ellipsoids. This finding that allows the understanding of the hydrodynamics of discrete irregularly-shaped particles and facilitates the estimation of the drag coefficient of different geometries based on the orientation of the particle respect to the fluid flow:

$$C_D = C_{D,min} + \left(C_{D,max} - C_{D,min}\right) \sin^2 \alpha \tag{3.1}$$

FIGURE 6 shows the interplay between the projected area versus the angle of polar rotation that follows the cyclic pattern. These values, plausible due to the geometry of the system, also allow to corroborate the periodicity of the behaviour of the drag coefficient of this type of geometries.

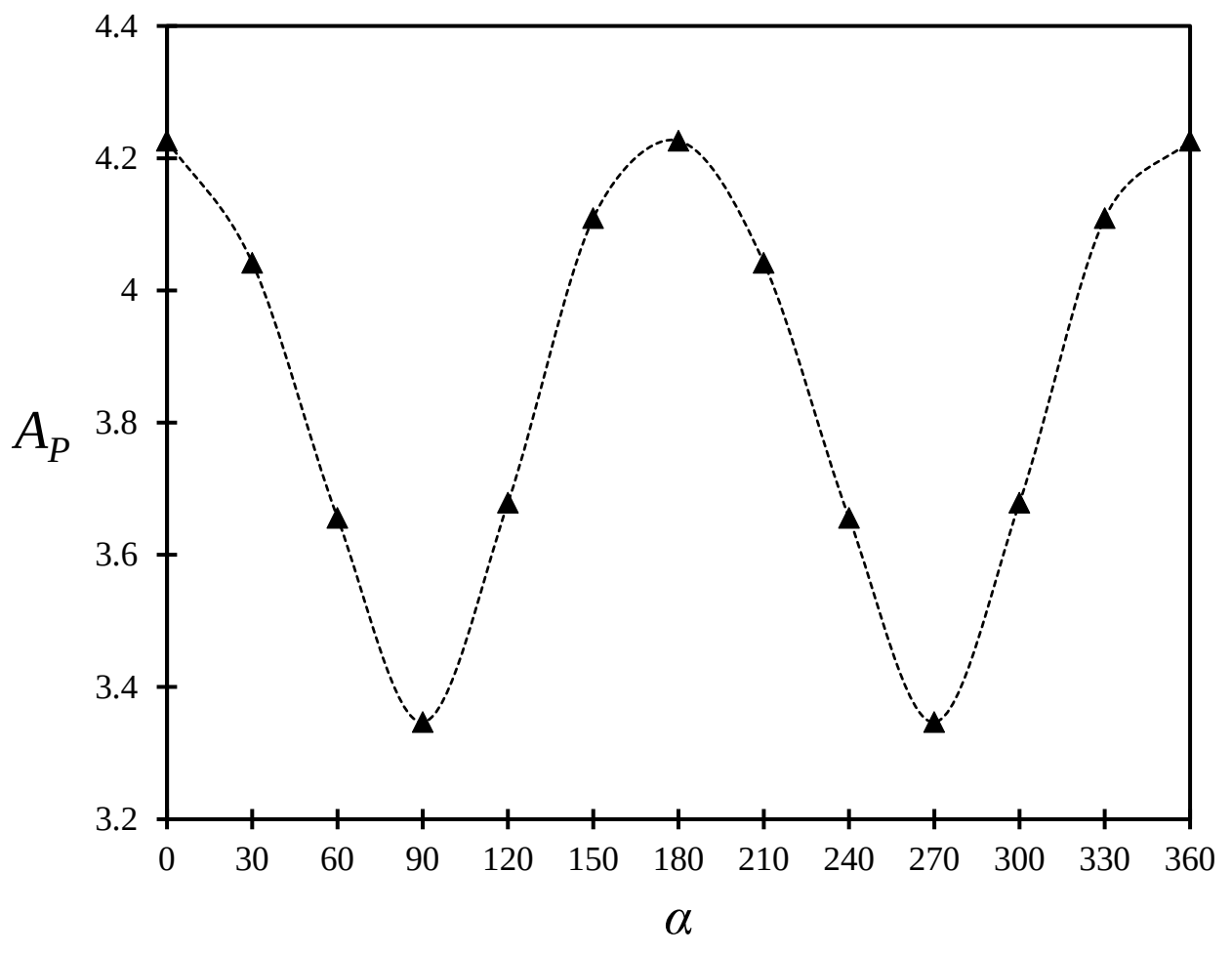

FIGURE 6. Projected equivalent area $A_P$ as a function of the rotation angle of incidence $\alpha$, for $\beta = 0$.

A statistical analysis has been developed considering thirty random combination values for both the polar rotation angle $\alpha$ and the azimuthal angle $\beta$. FIGURE 7 shows the distribution of the values of the drag coefficient for four different Reynolds numbers, both in the laminar regime and in the turbulent regime. In the following, we prove that the phenomenon follows a Weibull distribution (Weibull 1951), First, we fit such distribution to the data as shown in the figure.

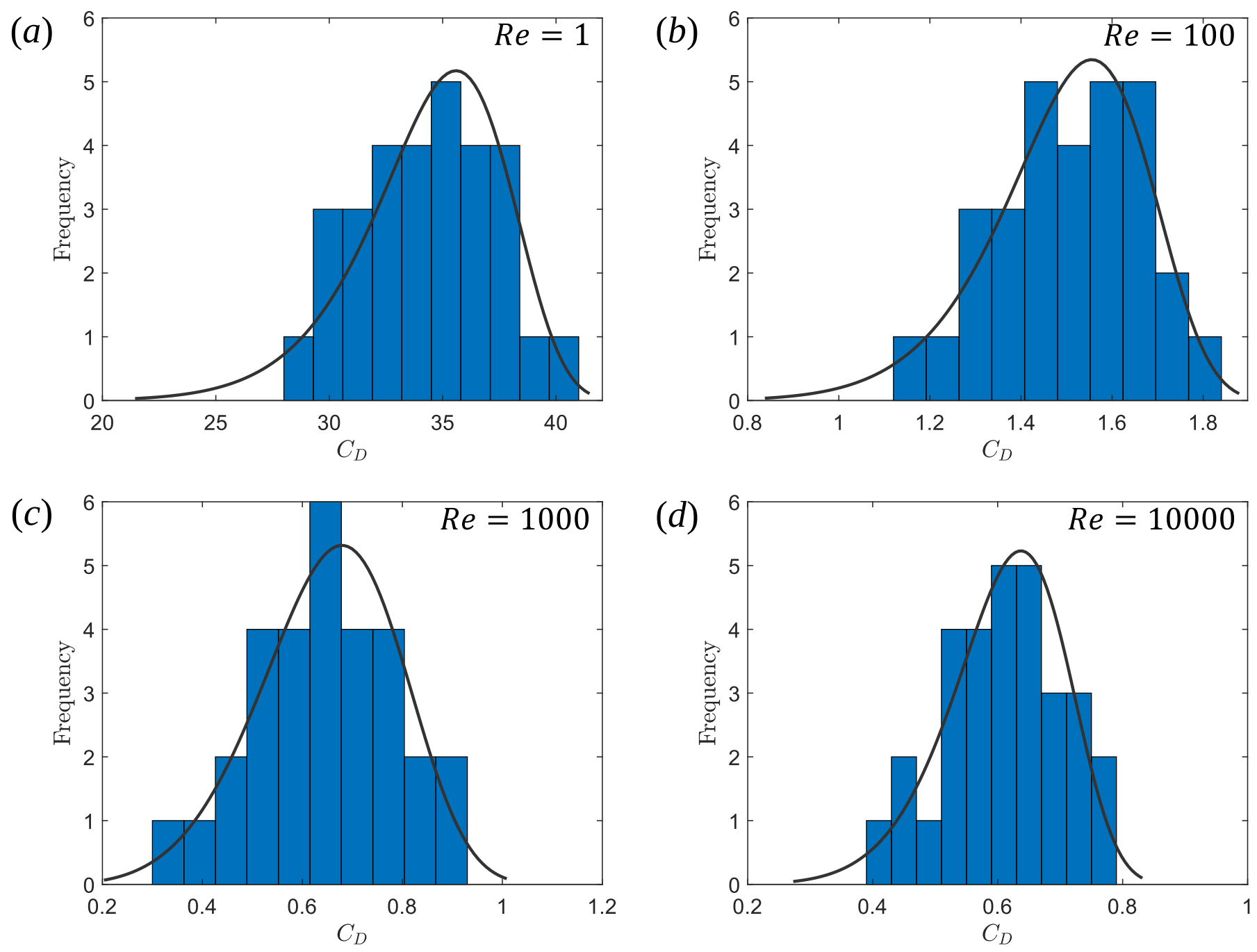

FIGURE 7. Weibull probability density distributions of drag coefficient for different random angles of incidence. *a*) $Re = 1$, *b*) $Re = 100$, *c*) $Re = 1000$ and *d*) $Re = 10000$.

Then, assuming a Weibullian behaviour of the drag coefficient, we can determine the probability of finding its value given a defined rotation, $P_S(\alpha, \beta)$, as:

$$P_S(\alpha, \beta) = \exp\left(-\frac{C_D}{C_{D,0}}\right)^m \tag{3.2}$$

where $C_D$ is the drag coefficient, $C_{D,0}$ is the characteristic drag coefficient and $m$ the Weibull modulus. Furthermore, the probability can be defined based on the number of $N$ simulations performed (in this case, 30 for each Reynolds number) in the mean rank position $i$:

$$P_S = 1 - \frac{i}{N+1} \tag{3.3}$$

Finally, equation 3.2 can be described in logarithmic terms as:

$$\ln[\ln(1/P_S)] = m \ln C_D - m \ln C_{D,0} \tag{3.4}$$

From equation 3.4, it is possible to find the parameters of the Weibull modulus $m$ as shown in FIGURE 8. The large $R^2$ ($> 0.975$) values show the fidelity of the Weibullian behaviour.

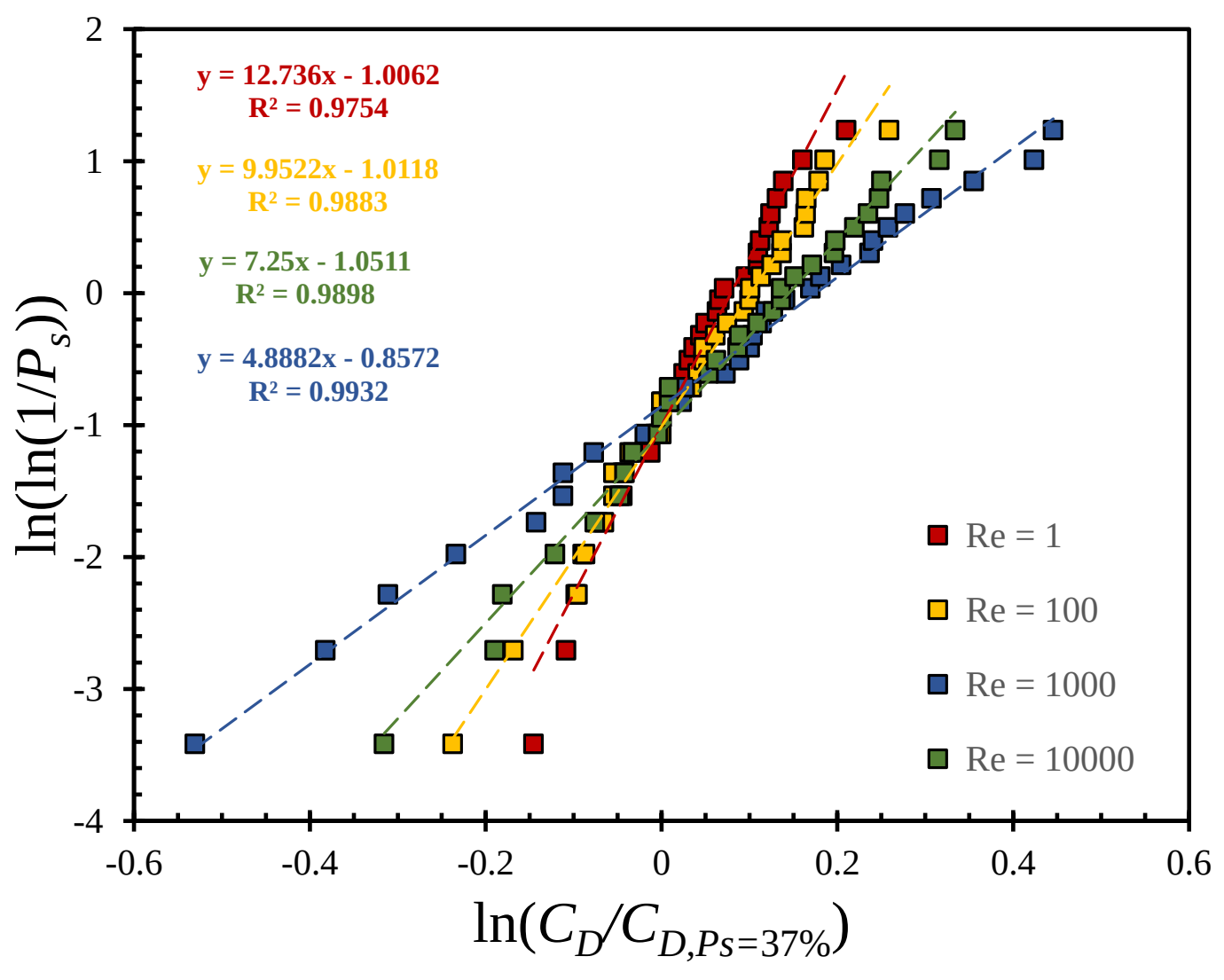

FIGURE 8. Weibull distribution for the 120 simulations obtained that describe the rotational dependence of the drag coefficient, for different $Re$.

### 3.3. *Boundary layer separation and turbulence*

One of the characteristics of external flows is the existence of a thin layer between the solid body and the fluid. This boundary layer varies depending on the flow regime and the surface of the solid body. In the case of smooth spheres, it is well studied that boundary layer separation occurs at approximately 80° (relative to flow direction) in the subcritical regime (prior to drag crisis) and 120° for the supercritical regime (Achenbach 1972). However, in irregularly-shaped particles more than one separation point is possible, and there is ambiguity in defining the separation angle, so we restrict ourselves here to show its existence in the turbulent regime. In FIGURE 9 we present the velocity field in the critical regime, at $Re = 100000$. The influence of the irregularity of the particle surface in the separation of the boundary layer can be observed, where defining its geometry precisely is complex, considering the flow patterns in it. Thus, in FIGURE 9*a* the pattern of the velocity streamlines is shown, where its increase occurs mainly at the poles of the particle, giving rise to the separation of the boundary layer. FIGURE 9*b* represents the velocity contours, highlighting the formation of turbulence and the chaotic pattern behind the particle.

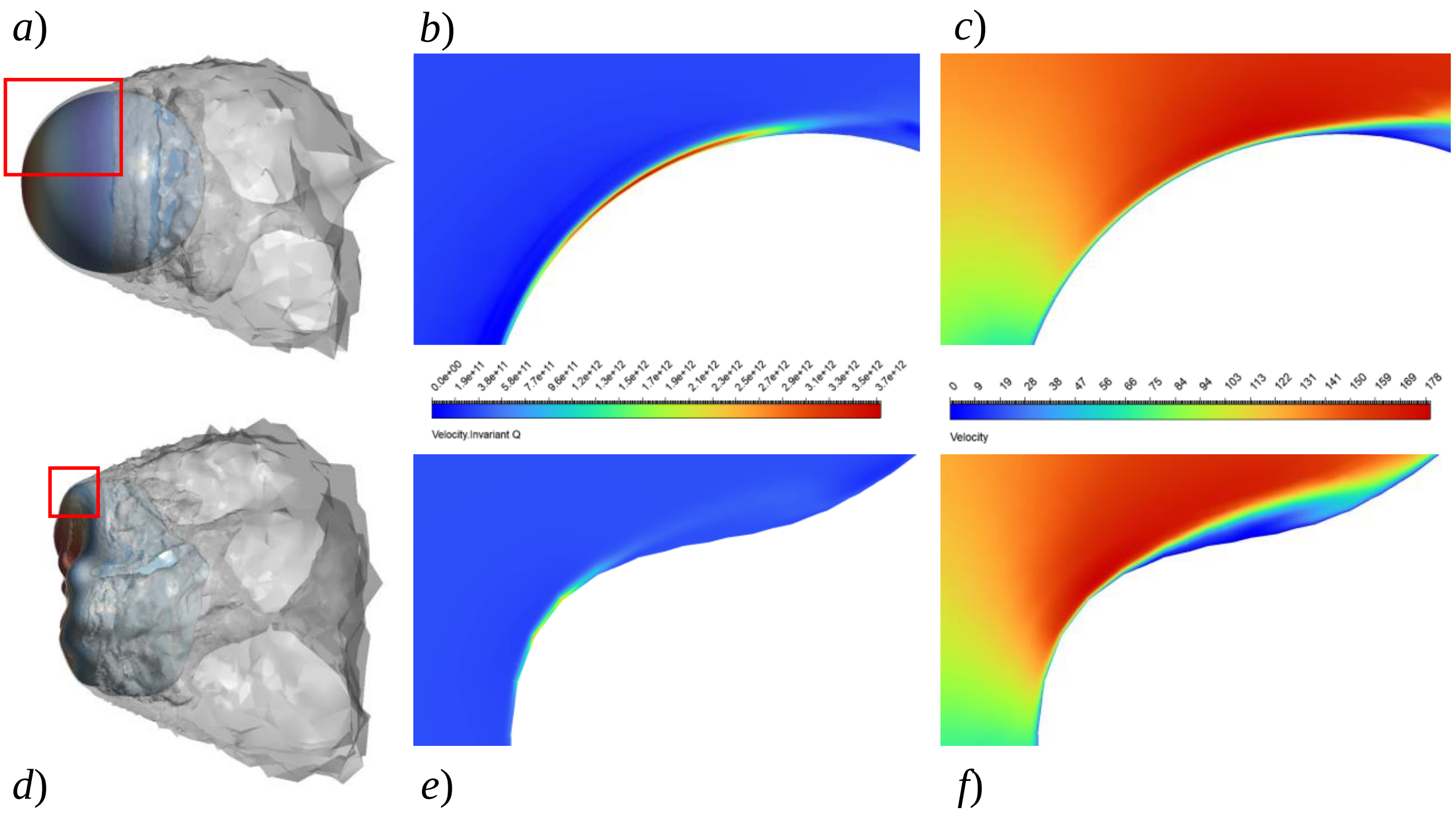

FIGURE 9. Turbulence patterns of the sphere (a, b and c) and the irregular particle (d, e and f) for $Re = 10000$. a) and d) represent the vortex isosurface around the sphere and the irregular particle, respectively. b) and e) correspond to Q-Criterion. c) and f) are the velocity fields on a meridional plane.

FIGURE 10*a* shows the pressure pattern at $Re = 10000$. The high-pressure points are concentrated on the front face that is normal to the flow, impacting the latter directly on the entire perpendicular area exposed to the fluid. FIGURE 10*b* shows the shear stresses for the same regime. Here, the points that experience the most shear coincide with the points of least pressure. This phenomenon occurs due to the characteristics of the boundary layer, which allows the flow lines to be displaced by "dragging" the fluid around the particle surface.

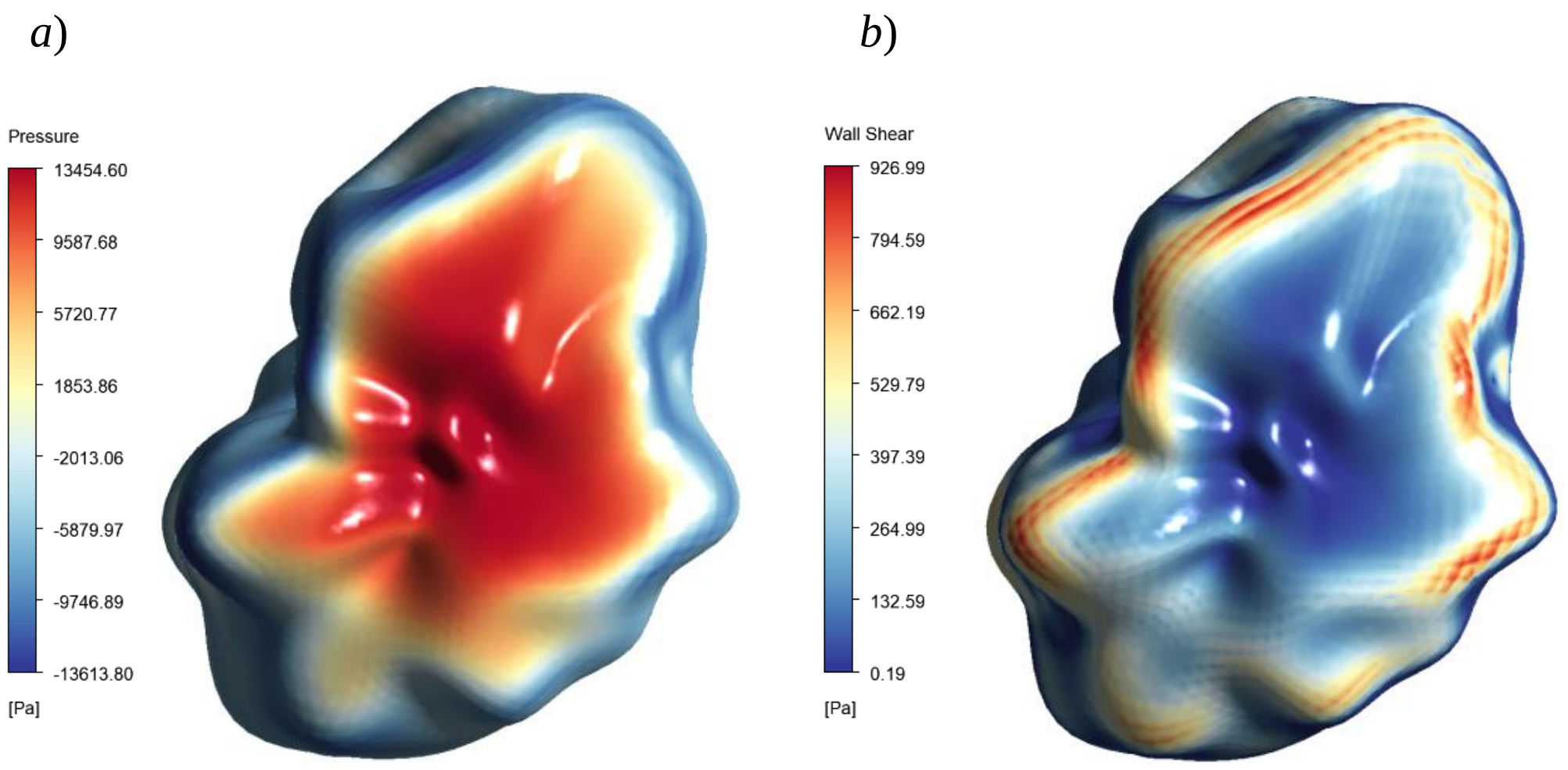

FIGURE 10. a) Pressure profile on the front surface of the flow. b) Shear profile on the front surface of the flow. Both figures correspond to turbulent flow in the critical regime $Re = 10000$.

### 3.4. *Influence of the geometric definition of the irregular shape*

Considering the complexity of defining the geometry of an irregularly shaped particle, the behaviour of the drag coefficient has been obtained considering three simple geometric definitions, which would modify the numerical value of the Reynolds number: projected area diameter $d_p = \sqrt{4A_p/\pi}$ (diameter of a circle that would have the same area projected $A_p$ in the normal direction of the fluid motion), area equivalent diameter $d_A = \sqrt{4A/\pi}$ (diameter of a sphere that would have the same surface area $A$ as the particle), and volume equivalent diameter $d_V = \sqrt[3]{6V/\pi}$ (diameter of a sphere that would have the same volume $V$ as the particle).

As a first glance, FIGURE 11 presents the drag coefficient as a function of the Reynolds number considering the three types of equivalent diameter definitions.

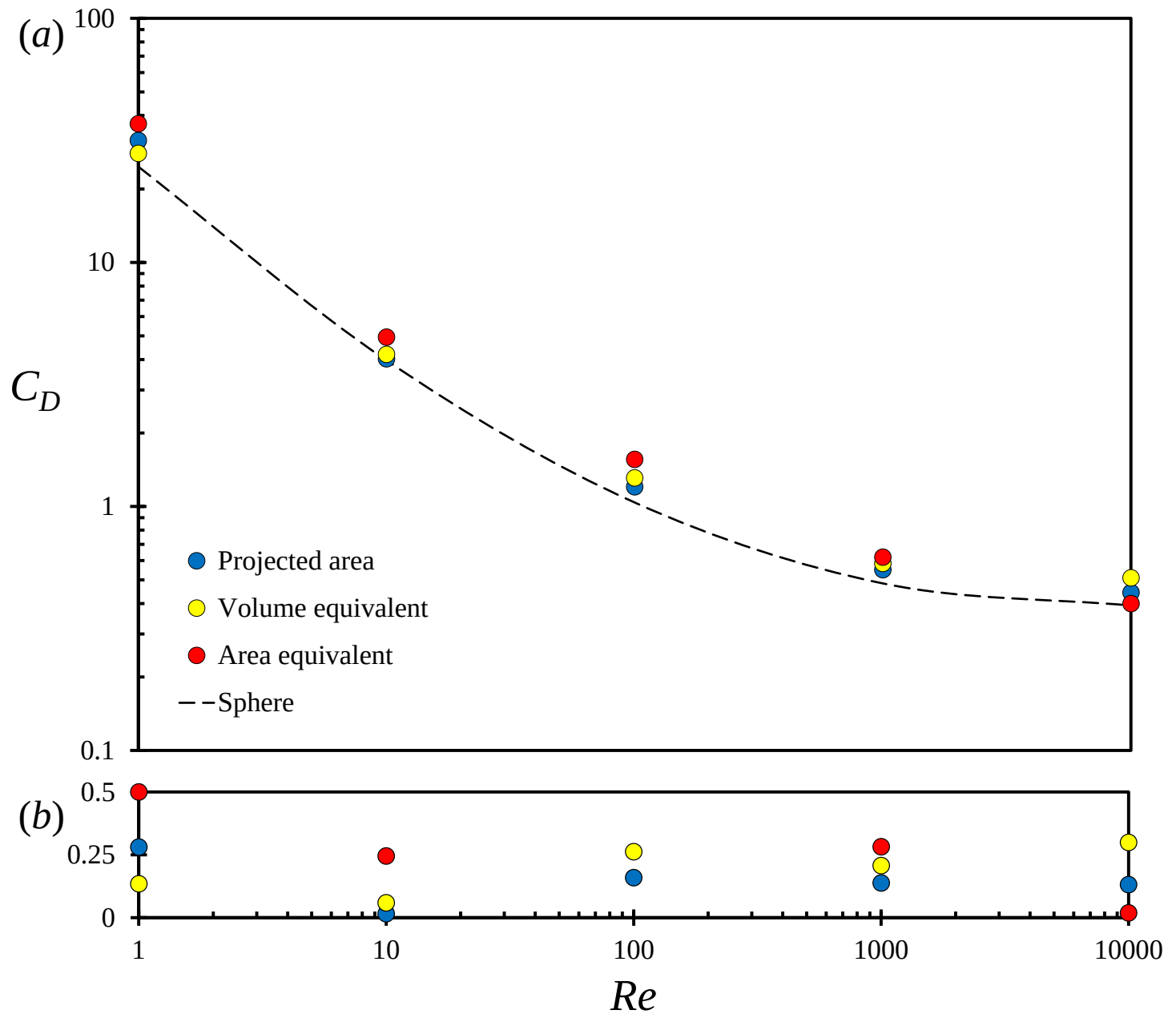

FIGURE 11. Drag coefficient $C_D$ in function of Reynolds number $Re$, for different equivalent diameter definitions.

It can be seen that the effect of considering different definitions of equivalent diameter does not seem to greatly alter the value of the drag coefficient. In this case, it is worth mentioning that modifying the definition of the equivalent diameter directly affects the obtaining of the Reynolds number, but not the drag coefficient, which depends geometrically on the projected area of the particle in the plane normal to the flow, for which its value in this case becomes dependent on the Reynolds number.

Finally, if we analyse the rotational dependence of the drag coefficient by modifying the definition of the equivalent diameter, we can observe in FIGURE 12*a* the variation in the results, without a well-defined pattern. However, when normalising the values, we find the pattern obtained previously. That is, regardless of the definition of equivalent diameter, they all fit a sine-squared interpolation (FIGURE 12*b*). These results have been compared with the behaviour obtained for an oblate spheroid at $Re = 100$ and justifies that the projected area diameter is the one that best fits the function, being the reason why it was considered throughout the study.

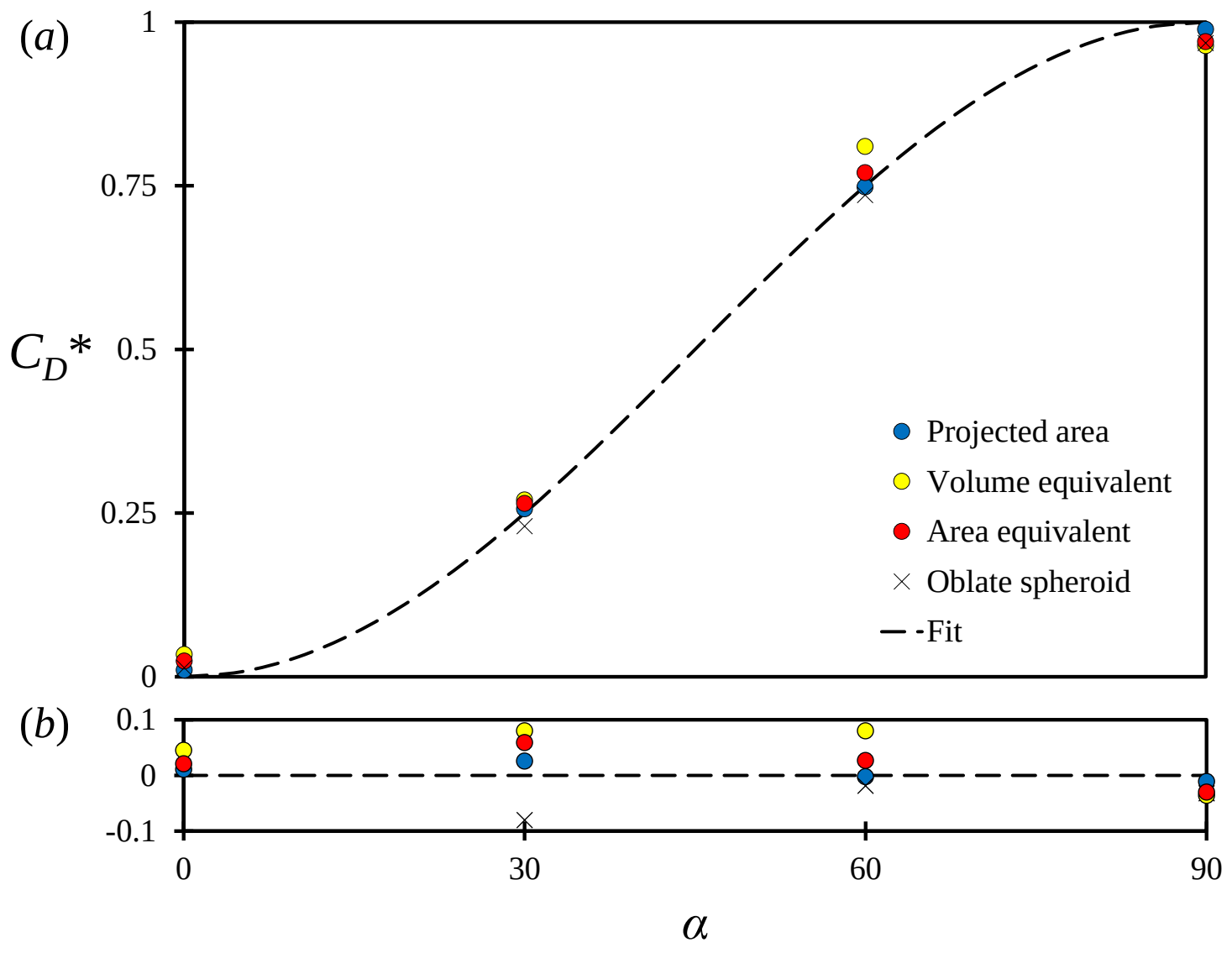

FIGURE 12. (*a*) Normalised drag coefficient $C_D^*$ as a function of the incidence angle $\alpha$ for different equivalent diameter definitions, compared with the behaviour of an oblate spheroid at $Re = 100$ (Sanjeevi & Padding 2017). (*b*) Percentage difference of the values of the normalised drag coefficient with respect to the sine-square interpolation, for $Re = 100$.

## 4. Conclusions

The drag coefficient of irregularly shaped particles in a wide range of $Re$ and flow incidence angles was studied. At high $Re$, we show the existence of the drag crisis for this type of geometries, a well-described and understood phenomenon for bodies of simple geometries. The angle of incidence shows a periodic behaviour of the drag coefficient with respect to the polar angle of incidence of the flow, and fits to a sine-square function. This also occurs even at very high Reynolds numbers (turbulence), thus extending the model previously described for spheroids, ellipsoids and smoothed shapes within the laminar regime. The value of the drag coefficient can be therefore estimated for any angle of rotation (around the same axis) knowing only two values, in this case, for an angle of incidence of 0° and 90°, using the suggested normalised definition of the drag coefficient. The separation of the boundary layer in irregular bodies is characterised by flow patterns that are not well defined due to the irregularity of the particle surface, but which relies directly on the regularity of the surface in front of the flow, inducing the formation of vortices and eddies past flow depending on the degree of separation of the boundary layer.

With this, we hope to inject interest in the study of the hydrodynamic behaviour of irregularly shaped particles, to extend its knowledge and encourage the analysis of highly turbulent flows around irregular bodies. This will also be useful to endless applications in geosciences and different disciplines of the engineering.

**Acknowledgements.** This work is part of the support granted by the National Scientific Research Agency of Chile - Grant for Doctoral Studies no. 62210003.

**Declaration of interests.** The authors report no conflict of interest.